\documentclass[conference]{IEEEtran}
\IEEEoverridecommandlockouts

\usepackage{graphicx}
\usepackage{textcomp}
\usepackage{makecell}
\usepackage[bottom]{footmisc}
\usepackage{multicol}
\usepackage{multirow}
\usepackage{balance}       % to better equalize the last page
\usepackage{graphics}      % for EPS, load graphicx instead 
\usepackage[T1]{fontenc}   % for umlauts and other diaeresis
\usepackage{booktabs}
\PassOptionsToPackage{hyphens}{url}
\usepackage{microtype}        % Improved Tracking and Kerning
\usepackage{ccicons}          % Cite your images correctly!

\usepackage{cite}

\usepackage[ampersand]{easylist}
\usepackage{tabularx}
\usepackage{rotating}
\usepackage[many]{tcolorbox}
\usepackage{soul}

\usepackage{colortbl}
\usepackage{verbatim}
\usepackage{array}
\usepackage[inline]{enumitem}
\usepackage[export]{adjustbox}
\usepackage{pifont}

\newcommand{\MyBox}[1]{\vspace{3mm}\noindent\framebox[\columnwidth][c]{\parbox[b]{0.95\columnwidth}{ #1 }}\vspace{3mm}}

\newcommand{\inlinequote}[1]{%\textcolor{NavyBlue}
{\textit{``#1''}}}
\newif\ifdraft
\draftfalse %hides boldifications and comments
\newcommand{\boldification}[1]{\ifdraft\indent ** \textbf{#1} **\\ \indent\else\relax\fi}

\makeatother

\begin{document}

%\def\BibTeX{{\rm B\kern-.05em{\sc i\kern-.025em b}\kern-.08em
%    T\kern-.1667em\lower.7ex\hbox{E}\kern-.125emX}}

%__________________ end packages and commands from paper__________

\title{Rules of Engagement: Why and How Companies Participate in OSS}

%\author{}%\IEEEauthorblockN{1\textsuperscript{st} Omitted for Anonymous Review}
%\IEEEauthorblockA{\textit{Omitted} \\
%email@email.com}
%}

\author{

\IEEEauthorblockN{Mariam Guizani}
\IEEEauthorblockA{\textit{EECS. Oregon State University} \\
Oregon, USA \\
guizanim@oregonstate.edu}
\and

\IEEEauthorblockN{Aileen Abril Castro-Guzman}
\IEEEauthorblockA{\textit{Oregon State University} \\
Oregon, USA \\
castroga@oregonstate.edu}
\and

\IEEEauthorblockN{Anita Sarma}
\IEEEauthorblockA{\textit{EECS Oregon State University} \\
Oregon, USA \\
anita.sarma@oregonstate.edu }

\and
\IEEEauthorblockN{Igor Steinmacher}
\IEEEauthorblockA{\textit{Northern Arizona University} \\
Arizona, USA \\
igor.steinmacher@nau.edu}

}

\maketitle

\begin{abstract}
Company engagement in open source (OSS) is now the new norm. From large technology companies to startups, companies are participating in the OSS ecosystem by open-sourcing their technology, sponsoring projects through funding or paid developer time. However, our understanding of the OSS ecosystem is rooted in the ``old world'' model where individual contributors sustain OSS projects. In this work, we create a more comprehensive understanding of the hybrid OSS landscape by investigating what motivates companies to contribute and how they contribute to OSS. %
We conducted interviews with 20 participants who have different roles (e.g., CEO, OSPO Lead, Ecosystem Strategist) at 17 different companies of different sizes from large companies (e.g. Microsoft, RedHat, Google, Spotify) to startups. 
Data from semi-structured interviews reveal that company motivations can be categorized into four levels (Founders' Vision, Reputation, Business Advantage, and Reciprocity) and companies participate through different mechanisms (e.g., Developers' Time, Mentoring Time, Advocacy \& Promotion Time), each of which tie to the different types of motivations. We hope our findings nudge more companies to participate in the OSS ecosystem, helping make it robust, diverse, and sustainable.
\end{abstract}

\begin{IEEEkeywords}
Open Source, OSS, companies in open source, motivations, diversity
\end{IEEEkeywords}

\section{Introduction}
\label{sec:intro}

%\boldification{The OSS has shifted where more and more companies are involved in it. Companies stopped viewing OSS as an ecosystem that was a threat to their existence and started seeing it an ecosystem of opportunity.}
%and Google's first release of Android in 2008, the OSS community

%\boldification{Companies' view of OSS has transitioned from a risky endeavor to an ecosystem of opportunity}

\boldification{***Open source transitioning from individual, decentralized participation to more and more companies participating. This has shaped the new OSS ecosystem: entanglement of individuals, and companies, big and small.***}
Open Source Software (OSS) is no longer a ``weekend warrior's endeavor"~\cite{robles2019twenty}. Over the last 20 years, the OSS ecosystem composition has changed drastically. OSS is now fundamental to company operations--not only for the code that they depend on, but also for their role in an ecosystem to which they actively contribute~\cite{robles2019twenty, fitzgerald2006transformation}. This is a paradigm shift from the early days when OSS was viewed as a threat that commoditized software to today where individuals and companies work symbiotically. About 77\% of respondents to the 2022 State of Open Source survey \cite{2022OSSSurvey} said they increased the use of OSS in their organizations over the last 12 months. In fact, major companies are now increasingly spearheading the development of state-of-the-art OSS technology (e.g., Kubernetes) \cite{casalicchio2020state}. 

%\boldification{Research has mainly focused on studying OSS from the perspective of the individual contributors, to understand their motivation, roles, and challenges}

\boldification{**Prior work has identified the hows and whys of individual participants.}
Research thus far, however, has largely investigated the OSS ecosystem from individual contributors' perspective: their experiences in OSS, their motivations to contribute~\cite{gerosa2021motivation, lee2017understanding, von2012carrots, hannebauer2016motivation, oreg2008exploring , roberts2006understanding, huang2021leaving}, the different ways to participate~\cite{jergensen2011onion, nakakoji2002evolution ,trinkenreich2020hidden,rebouccas2017does, schilling2012will, zhou2014will, steinmacher2021being}, and the challenges they face~\cite{steinmacher2015systematic, steinmacher2015social,guizani2021long ,lee2017understanding, huang2021leaving}. 

\boldification{***–which entails the old OSS ecosystem, our knowledge now has gaps in understanding the ecosystem, since we don’t know what role companies play.***}
Our understanding of how an OSS ecosystem works and how to sustain it is, therefore, modeled on the ``old world'' OSS ecosystem, and not the new hybrid reality, where ``every software company is an open source company'' \cite{eclipseOpenCollab}. 
We have little understanding of \textit{why} and \textit{how} companies get involved in OSS. 
It is important to understand what motivates companies to engage in OSS for multiple reasons. First, not doing so paints an incomplete picture of the OSS ecosystem, which is arguably one of the pillars of software engineering today~\cite{openSourceUseLinux, openSourceUse}. Second, without a clear picture of why companies participate in OSS, efforts to sustain OSS rely only on a subsection of contributors and are less likely to succeed. 
Third, highlighting the motivations and different mechanisms of engaging with OSS can help other companies recognize the benefits of being part of this hybrid OSS landscape, not only benefiting the company but also making the OSS ecosystem stronger. 

%A lack of understanding of companies' involvement with OSS paints an incomplete picture of the OSS ecosystem.  This, in turn, makes our research endeavors to sustain and maintain OSS incomplete and highly reliant on some OSS players and not others. As a consequence, researchers and OSS communities may end up putting efforts on solving a problem that leaves something out of the equation, which would require rework later. In order to put companies' participation in the loop, it is paramount to understand their motivations to be part of and the multi-faceted ways that they engage in OSS. This understanding paints a more complete OSS landscape that helps understand the relationship between different players and the OSS ecosystem.

\boldification{Prior work has mainly looked at licensing or how companies are doing inner sourcing.}
Past work that has investigated company involvement with OSS focused on
the legal perspective of companies' usage of OSS components~\cite{harutyunyan2019getting, harutyunyan2019industry, harutyunyan2018understanding}, their communication practices when interacting with OSS projects~\cite{butler2018investigation, butler2019company}, 
how adoption of OSS matches different company business models~\cite{hecker1999setting, munga2009adoption, spijkerman2018open}, 
or how companies are using the open source model internally (i.e., innersource) \cite{stol2014key, capraro2016inner, stol2014inner, edison2020inner, morgan2011exploring, capraro2018patch, carroll2017value, carroll2018examining}. These studies do not investigate the motivations of why companies contribute to OSS.

To the best of our knowledge, Linaker and Regnell's case study is the only work that has investigated company motivations---why and when companies decide to open source their products \cite{linaaker2020share}. They identified the objectives that led three organizations (a US media and technology organization, a European-based hardware electronics manufacturer, and the Swedish Public Employment Service) to open source their software. This work serves as a starting point for building a picture of why companies participate in OSS. 

In this paper, we aim to create a more comprehensive understanding of company motivations and mechanisms for engaging with the OSS ecosystem by asking:
\begin{enumerate}[leftmargin=0pt]
    \item[] \textbf{RQ1. }\textit{What motivates companies to contribute to OSS?}
    \item[] \textbf{RQ2. }\textit{How do companies contribute to OSS?}
\end{enumerate}

By creating an understanding of company motivations (\textbf{RQ1}) we aim to help projects provide an environment that is not only attractive to passionate individuals, but also to companies that can invest in and sustain the project in the long run.
Further, articulating companies' motivations to contribute to OSS can also nudge other companies to get more involved. By mapping the different ways companies contribute to OSS~(\textbf{RQ2}) we provide guidance to companies that are looking to engage in OSS. OSS projects can also reflect on the wide range of support companies provide and solicit support that aligns with their project needs. 
%Therefore, to guide these companies figure out how to contribute to OSS, we map the different ways companies engage with OSS by asking:

\boldification{*** [we did this] RQ’s explored through 20 interviews–with big and small companies, participants with different roles ***}
We answer our research questions through interviews with 20 participants from different companies, playing different roles related to OSS (e.g. Open Source Program Office (OSPO) Lead, OSPO Manager, CEO, Ecosystem Strategist) and working at companies of different sizes, including Microsoft, Google, RedHat, and Spotify. We then qualitatively analyzed the data through inductive coding to create a conceptual model of company motivations to engage with OSS and the different ways to do so.

%"you have these competing companies who collaborate everyday together and they have not signed anything."

%\boldification{Our paper's contributions include understanding why companies contribute to OSS, the different ways companies contribute, and discussion of lessons learned for a healthy/symbiotic relationship between companies and OSS}

\boldification{*** Our results can help:
Understand the impact of companies' participation on OSS 
Effectively mitigate companies' bus factors/ turnover
Provide projects with the knowledge to help them attract/ solicit the support they need from companies 
Effectively attract and retain companies to OSS 
Nudge companies to participate ***}

The main contributions of this paper are (1) a comprehensive model of company
motivations to contribute to OSS, (2) a conceptual model showing the multi-faceted ways that companies engage in OSS, linked to their motivations, and the benefiting entities, and (3) lessons learned from companies to foster a healthy OSS-company relationship. 
%\boldification{We hope our contributions help guide more companies to get involved in OSS in a ``symbiotic'' way-- as learned through our findings}
We hope that our insights on company motivations and contributions to OSS would encourage and provide guidance for more companies to engage in OSS, collaborate in the open, create better software and broaden the economic pie. The lessons we learned from this study will help both companies and OSS projects continue to foster a symbiotic relationship. %in which they sustain each other. % thus activating the ``virtuous cycle.''

\section{Research Method}
Because our study focuses on \textit{why} and \textit{how} companies contribute to OSS, we performed semi-structured interviews \cite{baltes2022sampling, ralph2020empirical} targeting interviewees with different roles that are knowledgeable about the company's involvement in OSS (e.g., OSPO Lead, CEO, Ecosystem Strategist) working at companies of different sizes (7 small, 1 medium, 9 large) \cite{businessSize} from startups to large technology companies such as Microsoft, Google, RedHat, and Spotify. The following two sections detail our data collection, analysis, and member-checking approach.  

\textbf{Recruitment.} We recruited 20 interview participants from 17 different companies of different sizes. We emailed and sent direct messages via Twitter to participants whose contact information was publicly available. Our selection criteria consisted of participants who were either in a leadership position (e.g., Founder, CEO) or in an OSS-related position (e.g., OSPO Lead, Ecosystem Strategist). We first conducted 12 interviews and used a snowball approach to recruit more interviewees. We conducted 8 additional interviews. Table \ref{table:participants} shows the participants, their roles, and their respective company's details. 

\textbf{Data collection and availability.} We arranged meeting dates and times according to each interviewee's availability. All interviews were conducted online. The first author conducted the interviews and started by introducing themselves and the project, and getting permission to record the interview. The interviews were 30 to 60 minutes long. After this, we thanked our participants and compensated them with a 50-dollar gift card as a token of appreciation. The data collection resulted in 20 interviews which were then transcribed. This is in line with the anthropology literature, which states that a set of 10-20 knowledgeable people is sufficient to uncover the core categories in any cultural domain or study of lived experience \cite{bernard2017research}. 
The interview transcripts are confidential as per our institutional IRB restrictions. However, to aid verifiability, we have included the interview guide, the codebook with example quotes, and code networks in our supplemental material \cite{suppdoc}. Also, Table \ref{tab:motivationTable} gives examples of how the findings mapped to the observations.

\textbf{Data Analysis.} 
We conducted and analyzed the data iteratively and we used ATLAS.ti\footnote{https://atlasti.com/} to perform the data analysis. Atlas.ti is a software for qualitative research that assists with analytic procedures by providing tagging capabilities and visualizations (mind maps) allowing the researcher to visually examine features and relationships in the “coded” texts (transcripts). We qualitatively analyzed the transcripts by inductively applying open coding, whereby we identified the motivations and contributions that each participant reported. We built post-formed codes as the analysis progressed and associated them to respective parts of the transcribed interview text. We reached code saturation after 13 interviews where no new codes emerged. However, we decided to finish the rest of the 7 interviews as they were already scheduled and made via personal introductions (snowball). Next, we grouped all the codes into categories using ATLAS.ti network. We held weekly meetings with three researchers experienced in the domain and in qualitative research to discuss and adjust codes and categories until we reached agreement. The first author presented and described each category and the researchers went through the codes and the quotes backing them. We discussed the grouped categories, merged categories into higher-level themes, discussed the relationship between them, and adjusted codes and categories until we reached a consensus. This process took 36 weeks. 
For instance, we merged the codes ``engineer-to-engineer communication'' and ``timely user feedback'' under the subcategory name ``closer channels''. When we found the same meaning for a concept that had been coded differently for more than one excerpt, we discussed it until we found the appropriate concept that represented all the coded excerpts. For instance, within the higher-level category ``business advantage,''  we merged the subcategory, ``avoiding technical debt'' with ``business dependency on OSS'' as the need to avoid technical debt stems from the fact that the business depends on OSS. 
In addition to the motivations and contributions mentioned by interviewees, we also collected lessons learned that interviewees shared with us. 

\textbf{Member checking.}
After analyzing all interviews, we performed member checking to evaluate the validity of our findings by emailing our participants an editable document with our findings on company motivations and ways to contribute to OSS. Eight participants (P1, P2, P6, P7, P11-13, P20) provided their feedback. All eight participants verified our findings and did so via the Google document as well as an email response. They all agreed with our results, and P1 and P20 provided only rewording suggestions. P20 suggested rewording ``joining'' dependencies' leadership and governance to ``aspiring to join'' as they felt these roles need to be earned and voted in by the community (see section \ref{sec:developertime}). P1 suggested rephrasing ``gaining user fidelity'' in the description of ``building verifiable trust'' to ``gaining community trust'' (see section \ref{sec:reputation}) which to them then might result in user fidelity. We addressed the feedback from both participants by specifically rewording the respective parts of the results using their suggestions.

% \begin{table}[bht]
% %\footnotesize
% \caption{Interview participants' role and company details}
% \resizebox{0.48\textwidth}{!}{
% \begin{tabular}{lllll}
% \toprule
% {\textbf{ID}} & {\textbf{Role}} & {\textbf{Company Size}} & {\textbf{Has an OSPO}}   &\textbf{Company}  \\\hline
% P1 & Director of Sales & Small & No & C1 \\
% P2 & CEO & Small & No  & C1\\
% P3 & OSPO Sr. Product Manager & Large & Yes & C2\\
% P4 & OSPO Lead & Large & Yes  & C3\\
% P5 & Ecosystem Strategist & Small & No & C4\\
% P6 & Software Engineer/Advocacy Lead & Medium & Yes & C5 \\
% P7 & OSPO Engineering Director & Small & Yes & C6 \\
% P8 & Founder and CEO & Small & No & C7 \\
% P9 & CTO & Small & No & C8\\
% P10 & OSPO Lead & Large & Yes  & C9 \\
% P11 & OSPO Lead & Large & Yes & C2 \\
% P12 & Sr. Ecosystem Strategist & Small & No & C10\\
% P13 & Ecosystem Strategist & Small & No & C11\\
% P14 & Community Architect and Leadership Head & Large & Yes & C12\\
% P15 & Community Manager & Large & Yes & C12\\
% P16 & OSPO Lead & Large & Yes & C13\\
% P17 & OSPO Sr. Program Manager & Large & Yes & C14\\
% P18 & OSPO Lead & Large & Yes & C15\\
% P19 & OSPO Sr. Program Manager/Strategist & Large & Yes & C16\\
% P20 & OSPO Program Manager & Large & Yes & C17\\
% \bottomrule

% \end{tabular}}
% \label{table:participants}

% \end{table}

%small: 8
%medium: 1
%large: 9

%ospo yes: 11

\begin{table}[bht]
%\footnotesize
\caption{Interview participants' role, company size and company}
\resizebox{0.48\textwidth}{!}{
\begin{tabular}{llll}
\toprule
{\textbf{ID}} & {\textbf{Role}} & {\textbf{Company Size}} & \textbf{Company}  \\\hline
P1 & Director of Sales & Small  & C1 \\
P2 & CEO & Small  & C1\\
P3 & OSPO Sr. Product Manager & Large  & C2\\
P4 & OSPO Lead & Large  & C3\\
P5 & Ecosystem Strategist & Small  & C4\\
P6 & Software Engineer/Advocacy Lead & Medium  & C5 \\
P7 & OSPO Engineering Director & Small  & C6 \\
P8 & Founder and CEO & Small  & C7 \\
P9 & CTO & Small & C8\\
P10 & OSPO Lead & Large   & C9 \\
P11 & OSPO Lead & Large  & C2 \\
P12 & Sr. Ecosystem Strategist & Small  & C10\\
P13 & Ecosystem Strategist & Small  & C11\\
P14 & Community Architect and Leadership Head & Large & C12\\
P15 & Community Manager & Large  & C12\\
P16 & OSPO Lead & Large  & C13\\
P17 & OSPO Sr. Program Manager & Large  & C14\\
P18 & OSPO Lead & Large  & C15\\
P19 & OSPO Sr. Program Manager/Strategist & Large  & C16\\
P20 & OSPO Program Manager & Large  & C17\\
\bottomrule

\end{tabular}}
\label{table:participants}

\end{table}

\section{Findings}
\label{sec:findings}

%paper about donating projects to OSS 
%\subsection{What does OSS mean for companies?}
%"So corporations have always been pretty much command and control with a little bit of marketplace dynamic. But OSS is about this network, the network effect. Why are corporations doing that? Like, that's this third modality of sharing information that's incredibly effective. Corporations are using this modality of hierarchical command control, which is incredibly ineffective. So maybe corporations need to learn and participate in OSS. And that's that sort of got me from anti-OSS to learning about network production models, to running an OSPO. And now being very involved in lots of things related to OSS. And then this conversation" [P10]
In this section, we describe our findings on what motivates companies to contribute to OSS and the different ways companies contribute to OSS.%, and some lessons learned. 
\subsection{RQ1: What motivates companies to contribute to OSS?}

According to our interviewees, company motivations to contribute to OSS fall under four top-level categories: \textit{Founder(s)' Vision, Reputation, Business Advantage, and Reciprocity} (see Table \ref{tab:motivationTable}). We will discuss what we found for each of these categories in the following sections.

%yes, we have some. So of course, you can claim to be an expert, if you contribute regularly to a project, yes, kind of probably. So that's one way of also getting this kind of trust, building a reputation in that sense. 

% 3:17 ¶ 73 in [P2] Daniel-Bitergia interview-D.docx
%we need to use other marketing efforts to be known and to be to be seen out there being being an open source company, I think, helps build like trustable branding.
% \usepackage{multirow}
\begin{table*}[htb]
%\centering
%\scriptsize
\caption{Company motivations to contribute to OSS}
\label{tab:motivationTable}
% \usepackage{multirow}
\resizebox{\textwidth}{!}{
\begin{tabular}{llll}
\hline
  \textbf{\begin{tabular}[c]{@{}c@{}} Category \end{tabular}} &
  \textbf{Subcategory} &
  \textbf{Participants} &
  \textbf{Exemplary Quotes} 
  \\\hline

   %____________________Founder's Ideology_________________________
 
  \multicolumn{2}{l}{\multirow{1}{*}{\begin{tabular}[l]{@{}l@{}} Founder(s)' Vision  \end{tabular}}}

%---------------------Founder(s)' ideology-----------------------  
   
  \begin{tabular}[l]{@{}l@{}} \end{tabular} 
  
  &

   \begin{tabular}[l]{@{}l@{}} P1, P2, P5, P7,\\ P9, P12 \end{tabular} 

&
 
 \begin{tabular}[l]{@{}l@{}} ``Why open source? One: it's ideological. The founders just believed that was the right thing to do'' (P1)
  \\
 % ``The four founders basically were avid contributors of open source, and they decided that's, that should be our DNA'' (P7)
    ``The reason why we started [company name] was to create an open source alternatives to large enterprise'' (P9)
  
  \\
  
%``[founder's name]'s vision and intent for the project in the first place ''[P5]

   \end{tabular} \\ 
  
%----------- Problem is open source in its essence----------
  % &
  %\begin{tabular}[l]{@{}l@{}} Problem is open source in it's essence \end{tabular} 
 % &
  
   %\begin{tabular}[l]{@{}l@{}}\\ P12, P7, P9 \end{tabular} 
  % & 
    
  %\begin{tabular}[l]{@{}l@{}}

   %\\ 
   %``so <company name> was built on top of open source. So without open source <company name> wouldn't exist ''[P7]
  % \end{tabular} \\ 
  \hline
%____________________Reputation_________________________
  \multirow{12}{*}{Reputation} &
  
%-----------VISIBILITY-------------
  \cellcolor{gray!15}\begin{tabular}[l]{@{}l@{}} Visibility \end{tabular} &
  
  \cellcolor{gray!15}\begin{tabular}[l]{@{}l@{}} P1-P9, P11, \\ P13-P18 \end{tabular} &
  
  \cellcolor{gray!15}\begin{tabular}[l]{@{}l@{}} ``Where open source is superior is it gives us superior visibility'' (P15)\\ 
  
  %``So it's clear that brings visibility and bragging rights'' (P9)
  
 % ``We decided to do is write a blog post... Well, a bunch of people started commenting on that. And it raised awareness of [company \\name] because we wrote it, but we didn't mention [company name], really, in that entire article... so it's actually a great marketing \\tool to talk about technology to use, even if it doesn't promote your, your own product'' [P8] \\
%  ``Having companies involved can also help you with... marketing, right?... that helps give visibility to some of these these projects as\\ well'' [P11] \\
  %``open sourcing these bits and pieces of whenever you can make it like tangible and like very transparent about \\ what was actually going on inside of the company. So that can that can help the deal of branding [P4]''
  ``If you actually came up with this standard, then you also get a lot of exposure'' (P4) %... And is again it kind of is like our desire to be able \\ to be a standardized piece of technology'' (P4)
 \end{tabular} \\

  &
  \begin{tabular}[l]{@{}l@{}} Building\\verifiable\\trust \end{tabular} &
  
  \begin{tabular}[l]{@{}l@{}} P1, P2, P5-P7,\\ P15, P17, P18 \end{tabular} &
  
  \begin{tabular}[l]{@{}l@{}}
  ``If we have a proprietary component...you have to trust the vendor to know whether or not that's being maintained...\\Whereas if it's an open source project,  I can watch the contribution history and I can know whether or not it's being maintained'' (P15)\\
  
  ``The biggest benefit from contributing is building trust'' (P18) %, and from trust comes influence. So you can actually influence the direction of a\\ project, especially if you're very dependent upon the project

  %``Because they were asking if we have some experts on X, Y, Z? \\ And then it's like, yes, we have some. So of course, you can claim to be an expert, if you contribute regularly to a project'' (P7)\\
  \end{tabular} \\

  &
  \cellcolor{gray!15}\begin{tabular}[l]{@{}l@{}} Networking \end{tabular} &
  
  \cellcolor{gray!15}\begin{tabular}[l]{@{}l@{}} P1, P2, P5-P8 \\P13 \end{tabular} &
  
  \cellcolor{gray!15}\begin{tabular}[l]{@{}l@{}}
  %``We've had two three customers... the biggest ones we've  had... And they decided to go with us, not that much because of the open\\ source product, but... the open source knowledge that we are there we are positioned there. So then we should hire these people\\ because they know'' [P2]\\
  ``Puts us in a position where we are we are surrounded by other principle aligned practitioners'' (P5)\\ 
  
  %``we try to contact as many people as we knew, and then asking them, \\you know to introduce to ask new people and so on. 
  ``So we were kind of trying to nurture our network of people and learning from them'' (P2)

   \end{tabular} \\
   &
  \begin{tabular}[l]{@{}l@{}} Attracting\\talent \end{tabular} &
  
  \begin{tabular}[l]{@{}l@{}} P3-P6, P8-P10, \\P13, P15 \end{tabular} &
  
  \begin{tabular}[l]{@{}l@{}}
  
  ``Open sourcing is one of the ways that that can make it much more interesting for outside candidates'' (P4) \\
  
  %``A lot of people decided to come here, be hired here, because the open source products'' (P6)\\ 
  
  ``It's good for recruitment...if my engineering tool is open source, and I need to hire new people, for the engineering team, \\the first group of people I look at is people who've made random contributions to that tool'' (P15)\end{tabular}\\

   &
  \cellcolor{gray!15}\begin{tabular}[l]{@{}l@{}} Fostering\\adoption \end{tabular} &

  \cellcolor{gray!15}\begin{tabular}[l]{@{}l@{}} P2, P4, P5, P8, \\P10, P11, P15,\\ P18 \end{tabular} &
  
  \cellcolor{gray!15}\begin{tabular}[l]{@{}l@{}}``It [open source] has been fundamental instrumental to our go-to market and our adoption'' (P8) % and just how we build our company'' [P8] \\ 
  \\
  
  ``They ended up eventually open sourcing theirs as well, because there wasn't any real adoption of it as a proprietary software'' (P15)\\
  
  %``And in fact, large tech companies compete with each other as to how much they can give away, and how much adoption they can get on\\ what they give away'' (P10)%So there's quite a competition between, oh, I'll open source something, you'll open source something they\\ both do about the same thing''[P10] \\
  
  \end{tabular} \\ 
 \hline

 \multirow{8}{*}{\begin{tabular}[l]{@{}l@{}} Business \\Advantage \end{tabular}} 

   &
   
  \begin{tabular}[l]{@{}l@{}} Business\\dependency\\on OSS \end{tabular} &
  
  \begin{tabular}[l]{@{}l@{}} P2-P5, P7, \\ P9-P12, P14- \\P17 \end{tabular} &
  
  \begin{tabular}[l]{@{}l@{}}
  %``We want these projects to be healthy, alive and well maintained, and sustaining. Because we're depending on them'' [P9]\\
  ``Most systems as [product name] is built on open source software, that is like the the basis of everything is open source dependencies. \\ Then our engineers also contribute to open source projects  that they depend on'' (P4)
\\
``So we have a product line called [product name], which is kind of our flagship like looking forward most strategic project \\that we have at [company name], and that's built entirely on top of [OSS dependency]'' (P11)
  \\
  %``There's a journey that companies seem to go on from like... thinking they don't use it, to realizing that they do... to a point where \\companies start realizing they need to give back ... Ultimately, they know at this point, that a healthy ecosystem means good\\ business for them, too'' [P12]
  \end{tabular} \\ 

  &
 \cellcolor{gray!15}\begin{tabular}[l]{@{}l@{}}Coopetition  \end{tabular} &
 
  \cellcolor{gray!15}\begin{tabular}[l]{@{}l@{}} P3, P7, P8, \\P11, P13-P15, \\P17-P19 \end{tabular} &
  
  \cellcolor{gray!15}\begin{tabular}[l]{@{}l@{}}%``it's an implicit resource sharing agreement between companies that are working on the same problem'' [P3]
  ``The benefit is open collaboration, you know, I don't have to be the only person trying to figure out how to solve a problem'' (P14)
  \\
  ``You need to get...multiple vendors involved so that you can a) like improve your product because you have to like be accepting \\ opinions from lots of different people and then b) you get more market share that way'' (P3)
  
  \end{tabular} \\

   &
  \begin{tabular}[l]{@{}l@{}} Closer\\channels  \end{tabular} &
  
    \begin{tabular}[l]{@{}l@{}} P1, P5, P7, \\P13-P15 \end{tabular} &
    
  \begin{tabular}[l]{@{}l@{}}  
  %``One is the network that you've built. Um, and that, you know, we've also talked about in terms of like, how actually, this, you know, \\concretely helps you, in situation like, Hey, you wanna solve a problem? Will you actually know who to talk to?'' [P13]
  
  %``direct feedback...That is a huge benefit... you can look into the contributors\\ file and see who contributed... want to talk to them directly? Go ahead and do it'' [P14]
 
 ``I want to be able to, you know, how are my customers using it? What are the challenges they are facing?...I don't want it to be, \\you know, only coming by feedback and then somebody collects all the feedback and then massages the feedback'' (P14)
% \\
%``If you actually know everyone in the community...the key people in the community building that software, you're going to solve in like 15\\ minutes'' (P13)
  
  \end{tabular} \\

   &
  \cellcolor{gray!15}\begin{tabular}[l]{@{}l@{}}Innovation \end{tabular} &
  
    \cellcolor{gray!15}\begin{tabular}[l]{@{}l@{}}P11, P13, P16, \\ P17, P19 \end{tabular} &
 
  \cellcolor{gray!15}\begin{tabular}[l]{@{}l@{}}%``Innovation is a big part of our of our open source story... because we can... innovate with the rest of the ecosystem around these \\ core, open source technologies'' [P11]
  
  ``I would say that the biggest benefit that we see is innovation, because we can only do so much on our own'' (P11)
  \\
  ``Innovation is the other one...So the more that we're able to collaborate with external communities...The better the product'' (P17)\end{tabular} \\\hline

  \multirow{1}{*}{\begin{tabular}[l]{@{}l@{}} Reciprocity  \end{tabular}}  

%-----------Sustain ecosystem-------------  
   &
  \begin{tabular}[l]{@{}l@{}}  \end{tabular} &
  
    \begin{tabular}[l]{@{}l@{}}P4, P6, P7, \\P12-P14, P16, \\P17 \end{tabular} &
 
  \begin{tabular}[l]{@{}l@{}}
  %``As much as... every company can, in its own capacity, everyone should be contributing back, everyone should be making sure that \\they are not just the receiving end'' [P7]
 
  ``From the perspective of being a...trustworthy, honest and sincere participant in open source that we want to be part of the ecosystem,\\ you know, giving back as much as possible'' (P17) \\
  
  ``We can also share, what do we do...it's sharing experiences between different groups and people learning from each other'' (P7)
  \\
  %``I think tech plays a huge role in society. And there are not a lot of people that understand the technology and how the technology\\ is really made... so I've essentially tried to move increasingly in the space where I could be an advocate of this and have been trying\\ to find my way through that'' [P13]
  \end{tabular} \\ \hline

\hline 
\end{tabular}}

\end{table*}

%---------Start of FOUNDERS VISION________________
\subsubsection{Founder(s)' Vision}
\label{sec:founders}
%The founder(s) vision is usually from the origin of smaller companies' participation in OSS (P1, P2, P7, P5, P9, P12).
Founder(s)' vision can stem from the founder(s)' ideology (P1, P2, P5, P7, P12) attributed to a \inlinequote{philosophical point of view at the very beginning (P2),} \inlinequote{the company culture (P5)} and to the fact that \inlinequote{all of the co-founders have a deep history in open source (P12).} 
%Founder(s)' vision stems from the founder(s)' ideology (P1, P2, P5), the nature of the problem the company is solving (P9), or a mix of both (P7, P12).

Still, the founder(s)' vision can also be attributed to the nature of the problem that the company is solving and its ties to OSS (P9, P7, P12). In some cases, the reason for founding the company is \inlinequote{to create an open source alternative to large enterprise resource planning software (P9)} which makes the ideology become part of the company's core. 

%the fact that the nature of the problem that the company is solving is tied to open source in its very essence. And the reason for founding the company in the first place is
%\inlinequote{to create an open source alternative to large enterprise resource planning software [P9]}. 

\MyBox{Companies contribute to OSS because of their \textbf{founder(s)' vision} and/or because the problem they are solving is tied to OSS in its very essence.}

%--------------------end of founders vision---------------

\subsubsection{Reputation} 
\label{sec:reputation}

Our findings show that investing in OSS can be compared to public relations. For example, P4 compares the motivation to participate in OSS to that of \inlinequote{tak[ing] bits and pieces of the engine room into the open... by like normal PR [Public Relations] conference talks or they could do it like by open sourcing some of these things.}

%`` to take bits and pieces of the engine room into the open and they can do this by like normal PR conference talks or they could do it like by open sourcing some of these things''. 

Our analysis revealed that companies focus on building \textit{reputation} in different ways, which we categorized as follows: \textsc{Visibility}, \textsc{Building verifiable trust}, \textsc{Networking}, \textsc{Attracting talent}, and \textsc{Fostering Adoption}. 

%%%%%%%%%%%%%%%%%%%%%%%%%%%%%%%%%%%%%%%%%%%%%%%%%%%%%%%%%%%%%%%
\boldification{no matter the size of the company, visibility is a driver to being in OSS.}
\textsc{Visibility.}
Our results show that participating in OSS brings visibility to companies of all sizes (P1, P2, P4-P9, P11, P13-P18). For instance, P2 explains \inlinequote{we wanted to be there [OSS project] in our case because this brings a lot of visibility.} Companies join OSS to cement their name to a domain, expand their brand, or receive credit as the creator of a technology that is an industry standard.

\boldification{This visibility helps companies build an understanding of what they do inside, which helps attach the company brand to the domain of expertise that it is known for.}

For example, visibility helps companies signal an understanding of  \inlinequote{what [is] actually going on inside (P4)} by showcasing their technical achievements (P1, P2, P4, P7), which helps cement the name of the company in their domain of expertise.

\boldification{While this is the case for some companies, the visibility capabilities of OSS can also help a companies transition to a different domain of expertise}

While this is the case for a number of companies, it does not end here. Companies can also leverage OSS to expand their brand beyond that for which they are known for (P4, P11). In particular, OSS can help companies expand their domain of expertise as explained by P11: \inlinequote{so people tend to think of [company name] as like the people who did [specific product]. And we've really moved beyond that.}

\boldification{Now visibility doesn't have to mean the same thing to all companies, it can mean becoming more international for growing companies and making an internal technology the \textit{de facto} technology and being seen as the founder}

{Visibility can also hold different meanings to different-sized companies. For small companies, visibility can mean \inlinequote{becom[ing] more international (P6).} Whereas, for large technology companies, standardizing a piece of technology is the ultimate visibility goal (P3, P4, P18) because \inlinequote{if you actually came up with this standard, then you also get a lot of exposure (P4).} A good example of this is how \inlinequote{[for example] open-sourcing React has provided more value for Facebook than leaving it to themselves ever could have. Because (a) they get contributions now on that library; and (b) they get the sort of street credit of the founders of React (P3--who does not work at Facebook).}}

%_________________________End visibility__________________

\textsc{Building verifiable trust.}
In our context, building verifiable trust was described as gaining users' trust and/ or gaining the communities' trust. OSS by virtue of its open character, enables companies to openly contribute to a project. Therefore, these companies--especially startups--see OSS as a place where they can showcase their expertise to their users to gain trust (P1, P2, P5, P7, P15, P18). In fact, openly contributing to a project allows users to see the companies’ contributions, their quality, and the rate at which they maintain the project. This differs from the proprietary software world, where users have to trust the company as to whether or not a product is being maintained. P15 explains these differences: \inlinequote{If we have a proprietary component..., you have to trust the vendor to know whether or not that's being maintained... Whereas if it's an open source project,  I can watch the contribution history and I can know whether or not it's being maintained (P15).} 

Gaining users' trust is particularly relevant when a company is at the forefront of a new technology or concept as explained by P2: \inlinequote{And they [users] say, well, I need someone, I need help with [name of concept]... So it's not that we know about [name of concept] it's that we are defining [name of concept] and for this, you can see that we have this pattern here about the maturity model.} This suggests that companies that are working on, or defining a new concept in the open, are more likely to be trusted and potentially hired by users, as users can directly see their contributions and decide on its quality.

While giving your users \inlinequote{open source value (P5)} builds verifiable trust, making it long-lasting requires going the extra mile--listening to and embracing users' feedback: \inlinequote{and what happens when you say, I'm giving you this, and then you deliver this, your users and your customers trust you. And they are more likely to stay with you (P5).}

It is not only startups who seek to build trust, we also found that for large technology companies (P14, P15, P17, P18) \inlinequote{the biggest benefit from contributing is building trust (P18).} However, this trust is specifically directed towards signaling a sense of \inlinequote{accountability to communities (P17)} and showing that \inlinequote{you are willing to sidestep what you think is best and make sure that decision makings include the community (P17).} That built trust is what then enables companies \inlinequote{to be at the table (P18)} and participate in shaping the direction of the projects that they depend on. 

In summary, different types of companies have different goals in creating trust; startups (mainly) contribute to OSS to gain users' trust and a user community, while larger technology companies contribute to OSS to gain the communities' trust.

\textsc{Networking.} Participants described networking as socializing with the OSS community, other companies, and/or being in the same space as big-name companies. This allows companies to open the lines of communication for collaborations and partnerships. Our analysis shows that networking can look different depending on the company's goal. Most small companies found networking to be a driver for OSS participation. 
Some of these companies saw networking as an opportunity to socialize with peer companies and organizations and get to know people in the same space. For instance, P5 depicted networking as being in \inlinequote{that same ecosystem as these other like-minded organizations and practitioners (P5)} and P7 described networking as \inlinequote{socializing on that level of getting to know people, talk to people (P7).}

Other companies saw networking as a way to be in the same space as big-name companies (P1, P2, P6). For example, P6 describes networking for their company as a way to \inlinequote{bring more people to pay attention...and engage our brand to be related with huge companies (P6).} Other companies emphasized networking as a way to ease collaborations. For example, P8 mentioned how networking made
\inlinequote{collaboration and partnerships become six orders of magnitude easier (P8).}

%as a motivation to contribute also looks different depending on the company's goal
%, with growing companies participating in OSS to be in the same space as big-name companies.

%which leads us to the next section, stay tuned ...

%\MyBox{\textbf{Networking:} Growing companies participate in OSS to be in the same space as big name companies, opening the lines of communications for collaboration and partnerships.}

%_________________ATTRACTING TALENT___________
\textsc{Attracting talent.}
Our results show that companies participate in OSS to be seen as a ``cool company'' to work at, hire talents that are familiar with their tech stack, and bypass the interview and onboarding process. Participant P3 shared that companies that contribute to OSS \inlinequote{get viewed as, like a cooler place to work (P3).} P4 explains that \inlinequote{open sourcing is one of the ways that can make it [company] much more interesting for outside candidates (P4)}
and signals to developers that, by joining such a company, they too can work on ``cool technology'' that has a lot of traction in the world (P3, P4, P8, P10, P13). 

Our findings suggest that such reputation is highly beneficial for companies, which can then hire some great people since, as P10 points out, \inlinequote{people who code in public are better coders (P10).} Not only that, but companies also \inlinequote{get an entire class of engineers, who are now familiar with the tech stack that [company] uses (P4)} thus bypassing the interview and training process (P3, P4, P6, P10). 

This is particularly relevant for large technology companies that have open-sourced a technology that has now become the industry standard. For instance, P4 (who does not work at Google) mentioned that \inlinequote{any engineer that Google hires now will know Kubernetes.} 

%So this mindset also, cause many companies that didn't go open source to lose talent and don't retain talent. So this works also as a way to retain talent, like, Hey, we are not just a boring company in the financial systems, I don't know, for instance, like, we are a tech company, and we are promoting open source and we really want you to create things and linear space to do so. Right? And also to like your Yeah, like your personal brand gets in the pilot for for your future career.[P19]

%Small companies also reap the benefits of open source contributions \inlinequote{to hire some of the very best in the field [P9]} even though they \inlinequote{don't have the ability to pay the same [P9]} as larger technology companies do. 

%\MyBox{\textbf{Attracting talent:} companies join OSS to be seen as a ``cool company'' to work, to hire talents familiar with their tech stack, and bypass the interview and onboarding process.}

\textsc{Fostering adoption} in our context refers to a company’s technology being widely adopted. Fostering adoption can be an integral part of building a company's reputation, so much so that the company \inlinequote{would rather focus on broader adoption than just you could say having our name on it (P4).} % its virtue of being open to anyone, OSS helps foster adoption. 

For smaller companies, and as described by P2, \inlinequote{the best way to get your technology adopted is by open-sourcing this (P2).} Our results suggest that open-sourcing technology not only provides a readily available alternative to proprietary software but levels the playing field for startups to compete with larger companies.

For larger companies, fostering adoption can look like, as described by P11, \inlinequote{not locking them [enterprise customers] into a proprietary technology} and giving them the option to deploy the same solution using a different vendor. Our findings show, that while
the open character of OSS helps foster adoption, some companies do not stop at the mere act of open-sourcing a product. Such companies \inlinequote{strategically give that [project] to a foundation to grow in public, so that more people use a technology (P10)} (P4, P11, P18). Our findings highlight that allowing a project to grow under neutral governing bodies gives additional impetus to wider adoption.

\MyBox{Contributing to OSS helps cement \textbf{reputation}. It gives visibility to smaller companies, allowing them to play in the same space as big-name companies. Larger companies benefit by developing  deeper trust with the community signaling good citizenship, which in turn helps in recruiting.}

%%%%%%%%%%%%%%%%%%%%%%%%%%%%%%%%%%%%%%%%%%%%%%%%%%%%%%%%%%%%%%%

%______________________START OF BUSINESS ADVANTAGE_______

%\textsc{Engineering need.} The business advantage journey starts with the engineering need: \inlinequote{thinking they don't use it, to realizing that they do... (P12).}

%OSS is home of \inlinequote{some of the greatest technologies, pieces of technology, around the world (P2).} With such realization companies start their OSS journey by \inlinequote{consum[ing] open source, that's the, the most obvious one, and like 99\% of the companies in the world do that (P7).} By consuming quality and readily available solutions, companies are able to \inlinequote{spin up this [product name] product line in a relatively short period of time (P11).} This--in addition to OSS's independence from licensing and certification needs (P7, P13, P15)--makes OSS solutions more attractive where companies \inlinequote{look to use open source whenever we can, for our own stuff, as well as open standards (P8).}

%\MyBox{\textbf{Engineering need}: companies consume OSS because it hosts high-quality software, it is independent of licensing/ certification needs and helps build quality software rapidly.} 

%---------------business depends on oss-------------------
\subsubsection{Business Advantage}
\label{sec:business}
Over the years, companies' mindset has shifted in a way that now sees OSS as an ecosystem of business opportunities rather than a risky endeavor that commoditizes their products \cite{robles2019twenty}: % This is clearly reported by one of our participants: 

\inlinequote{There's a journey that companies seem to go on from like...thinking they don't use it, to realizing that they do...to a point where companies start realizing they need to give back...Ultimately, they know at this point, that a healthy ecosystem means good business for them, too (P12).}

The Business Advantage motivation is divided into four different subcategories. While companies are driven by a direct business need to contribute to OSS, other motivations (Closer channels, Coopetition, and Innovation) also play a role when it comes to companies' involvement in OSS. We detail these motivations in the next subsections.

\textsc{Business dependency on OSS,} refers to cases where
throughout the companies' journey in OSS, their business becomes dependent on OSS \inlinequote{...to a point where companies start realizing they need to give back because they see that it actually helps them strategically (P12).}

Because the business depends on OSS projects, companies are motivated to contribute to keeping their dependencies healthy (P2-P5, P7, P9-P12, P14-P17). And just like an engine is critical to a car manufacturer: \inlinequote{if something happens to that engine...that works for Ford, right? Or, even worse, it makes it so that Ford can't use the engine the way they were using it before (P3).} Our results show that business dependency on OSS relates to ensuring that the OSS projects a company depends on remain healthy and that the leadership of these projects is not monopolized.

\boldification{Keep dependency healthy}
In fact, as described by P9, companies \inlinequote{want these projects [dependencies] to be healthy, alive and well maintained, and sustaining (P9).} To help with sustainability, companies contribute financially (P4) as well as upstream fixes (P17), which in return prevents carrying any technical debt internally (P11).
 
\boldification{software monopoly} 
The technical soundness of a project dependency is not the only aspect of its health. In fact, P7 explains that \inlinequote{part of the healthiness of the project is also making sure that there are different ideas, opinions, and points of view. It's really easy to fall to a kind of monopolistic way on the open source projects (P7).} Our findings suggest that participating in governance and leadership (P3, P7, P15, P11) helps ensure a shared roadmap. Our findings also highlight that, a shared leadership flows both ways. It cannot just thrive on companies' readiness to get involved in governance roles, it also relies on the project having an open leadership and governance that welcomes participation. When the project in question is led by a company and this company realizes that
 \inlinequote{we cannot be the only ones backing this thing up (P7),} they start seeking \inlinequote{a backing up of different companies (P7)} which in turn becomes a stepping stone for coopetition, which we detail in the next section.

%\MyBox{\textbf{Business dependency on OSS:} companies want to ensure that the OSS projects they depend on are healthy, and the leadership of these projects is not monopolized.} 

%___-----------------Coopetition_____________________
\textsc{Coopetition.}``Coopetition is the act of cooperation between competing companies''\cite{coopetitionDef}. Our findings suggest that in OSS companies can work together on the hard technical problem they all face and compete on other features around the problem. %\inlinequote{all these little things around the [problem] that make a [product] worth buying (P3).}

Our results suggest that coopetition yields more productivity where (1) workload is reduced and (2) products are of higher quality. The workload is reduced (P3, P8, P13-P15) as collaborating helps \inlinequote{do the hard work only once (P3)} and provides a form of resource sharing where \inlinequote{20 engineers will go further working on a project with 80 other engineers from four different companies (P3).}

We also found that the benefits of coopetition are not solely related to productivity. Such practice enables a bigger market share (P3, P15), which in return provides a much larger user base, one that is now empowered to migrate to different service providers that use the same core technology. 

Still, according to P3 and P11, the products end up having higher quality 
because \inlinequote{pulling in people who work from other companies who will use things differently...they'll have different skills...they'll have different ways of thinking (P11).}

%\MyBox{\textbf{Coopetition:} companies participate in open source to collaborate with other companies on the harder problems, share resources, build a better product, and have a bigger market share.} 

\textsc{Closer channels}
in our context, relates to more direct communications, such as direct user feedback and, engineer-to-engineer communication that promotes timely fixes.
OSS is, in fact, a no-middleman land that provides closer channels and engineer-to-engineer communication (P5, P7, P13-P15, P17). The closer channels can take the form of raw users' feedback or customers' collaborations (P5, P14) that is then, as described by P5, \inlinequote{ingest[ed] back into our product delivery process.} 

Closer channels also mean bypassing the company-to-company communication protocols %where \inlinequote{you cannot reach out to the developer directly (P14),} 
and making engineer-to-engineer communication the new norm (P13, P14). As described by P13, \inlinequote{it's fairly easy to waste a week on an incredibly dumb bug, that if you actually know everyone in the community... you're going to solve in like, 15 minutes (P13).}

%\MyBox{\textbf{Closer channels:} companies participate in OSS because it brings the channels of communication closer, such as readily user feedof mentori engineer-to-engineer communication that promotes timely fixes.} 

\textsc{Innovation.} OSS is, by virtue of its character, a space for innovation that provides a business advantage and drives companies' participation (P11, P13, P16, P17, P19). Our findings suggest that innovation takes the form of more perspectives, more user stories, and diverse ideas that shape the company's products (P11, P17, P19). This innovation boost then makes it more likely \inlinequote{that it [product] will solve problems for more people and not just those 12 extremely bright people in the room (P17).}

\boldification{Getting familiar with projects interesting in the future}
For \textit{avant-garde} companies, innovation can also take the form of getting familiar with projects that can be strategic in the future. For instance, P11 explains: \inlinequote{one of the things that we try to do is contribute to projects that we think are going to be strategic for [company name] in the future, but that aren't necessarily strategic for us right now (P11).}

%\MyBox{\textbf{Innovation:} companies participate in OSS because it fosters collaboration, yields new perspectives, and gives companies access to potential strategic projects in the future.} 

\MyBox{Companies start to use OSS as a \textbf{business advantage}, but then their journey evolves to contributing to maintain the tech they are dependent on, and they stay since participating in OSS allows them to cooperate with their competition and innovate quicker.}

%-----------Start of RECIPROCITY____________________

\subsubsection{Reciprocity} Companies sometimes want to give back to the OSS ecosystem by contributing to the projects they use or other projects to share knowledge and experiences for the greater good of OSS (P2-P5, P7, P9-P12, P14-P17).
%
%\textcolor{red}{
%Reciprocity can take the form of simply giving back to the OSS ecosystem that has given so much}. 
Our findings suggest that reciprocity stems from the realization that companies have extensively used open source and need to give back. For instance, P7 shared \inlinequote{we took lots of things from open source.} P9 explains that if \inlinequote{there's no reciprocity in some ways, even small, then you're missing the point entirely.}

In contrast with the contribution driven by the business depending on OSS (see section \ref{sec:business}), reciprocal contributions are ones that do not necessarily have a direct impact on the company but are rather, as P9 puts it, \inlinequote{the right thing to do.}

%\inlinequote{benefit for us there. I mean, there's no direct benefits in having the [project name] cleaned up, it was more we thought it was the right thing to do [P9]}.

Reciprocity can take the form of mentoring (P1, P2, P9, P12, P14, P17, P20),  \inlinequote{sharing experiences between different groups (P7)} (P6, P13, P14) and advocating OSS by \inlinequote{spreading the word. And letting everybody know that contributing to open source is good. It's good for you and your company (P6).}

\MyBox{Companies want to \textbf{reciprocate} by contributing back to the projects they use and to other projects for the greater good of OSS.}

%so there is no direct benefit for us there. I mean, there's no direct benefits in having the Linux kernel cleaned up, it was more we thought it was the right thing to do. It was pure pro bono investments, syncing the work we do with these tools. In the end, we believe that if more people use open source, more people will notice the service and products we provide. And, and and that will be beneficial to everyone in the long run. [P9]

The four motivation themes (RQ1) are what drive company contributions. To further understand the companies' engagement, in the following section, we link these motivations to the multi-faceted ways companies contribute to OSS, highlighting the entities that benefit from such involvement (see Figure~\ref{fig:contribute}).

\subsection{RQ2: How do companies contribute to OSS?}

\begin{figure}[!ht]
\centering
     \includegraphics[width=0.52\textwidth]{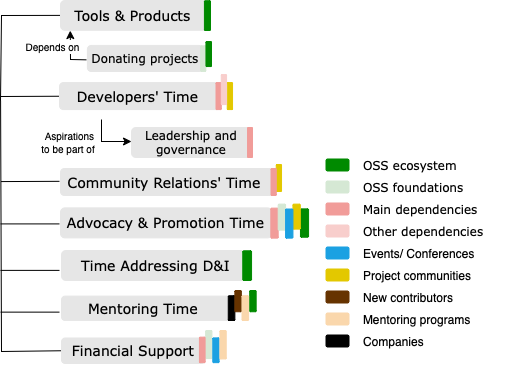}
    \caption{Company contributions and their beneficiaries}
    \label{fig:contribute}
\end{figure}

%\subsubsection{Contribute to open source}
When it comes to contributing to OSS, there is a number of ways to participate. We identified seven categories of contributions across companies: open sourcing \textsc{Tools \& Products}, \textsc{Developers' Time}, \textsc{Community 
Relations' Time},  \textsc{Advocacy \& Promotion Time}, \textsc{Time Addressing D\&I}, \textsc{Mentoring Time} and \textsc{Financial Support} (see Figure \ref{fig:contribute}).

%-------------SOFTWARE/tools and products----------------
\subsubsection{Tools \& Products} %Three of the four motivations that we identified as drivers for companies' participation in OSS (Founder(s)' vision, Business advantage, Reputation--Table \ref{tab:motivationTable}) point to open sourcing internal tools and products as a way to contribute to OSS (see Figure \ref{fig:motivation}).

In our interviews, we could observe that companies contribute by open sourcing \inlinequote{platforms and frameworks (P11),} \inlinequote{tools (P1),} \inlinequote{internal code (P4),} \inlinequote{[product name] community editions (P11),} \inlinequote{SDKs (P12)} and \inlinequote{those things in between, those connectors (P7).} Contributing internal products to OSS is a win-win for both the companies and the OSS ecosystem. 

Our results suggest that the internal tool, when open-sourced, benefits the whole OSS ecosystem (see Figure \ref{fig:contribute}) by enriching it with active projects that are backed up by one or more companies. For the company, the open-sourced tools provide a business advantage. For example, P7 describes such business advantage as \inlinequote{mak[ing] things easier for everyone using [company product]} and  \inlinequote{connect[ing] services that [company] does.}

Companies also reap the reputation benefit of open sourcing a novel solution, \inlinequote{a thing that does not exist} as P7 explains. 
The reputation benefit is magnified when such open-sourced tool is successful and widely adopted. For example, P4 explains \inlinequote{So [product name] is the thing that has the most success, and it's been adopted by a lot of companies... it's very impressive.}

Our results also suggest that when the question becomes \inlinequote{how much adoption they can get on what they give away (P10),} companies strategically donate their projects to an OSS foundation, signaling neutral governance. This, in return, helps sustain OSS foundations and the OSS ecosystem as a whole (see Figure \ref{fig:contribute}). 

Still, when the donated project becomes an industry standard, the company starts to be seen as the leader in that space which increases the company's visibility and reputation. For instance, P4 explains \inlinequote{because if it [OSS project] becomes an industry standard, then you gain all of the benefits from being part of the standards. And if you actually came up with this standard, then you also get a lot of exposure.}

%-----------------end of SOFTWARE------------------------------

%------------TIME and LEADERSHIP/GOVERNANCE---------------------

%\textsc{Time:} 90\% of companies depend on open source code \cite{openSourceUse} and the companies represented by our interviewees were no exception. 

When \inlinequote{the basis of everything is open source dependencies (P4)} contributing time is companies' first response. This takes the form of developers' time (P4, P6, P7-P20), community relations' time (P1, P5-P10, P11, P14, P15, P20), [time] on advocacy \& promotion (P1, P5-P9, P11-P16), [time] addressing D\&I (P5, P6, P13, P17, P19, P20) and [time] on mentoring (P1, P2, P9, P12, P14, P17, P20).

\subsubsection{Developers' Time}
\label{sec:developertime}
Our results show that contributing time often starts in an informal manner, led by developers. For instance, P8 explains:  \inlinequote {In the process of using, we tend to contribute to things we do use.} More specifically, developers' time can take the form of \textit{ad hoc} contributions such as fixing bugs (P9, P10, P12, P13) or adding a feature (P4, P8, P10, P11), where engineers \inlinequote{contribute where they see a need...[as a] scratching your own itch kind of thing (P4)}.% as a \inlinequote{(P4).}

When there are \inlinequote{300 things that depend on this specific dependency (P4),} contributing time in an \textit{ad hoc} way is no longer a viable option. Companies step to the next level, investing developers' time in a more systematic way. This can take the form of assigning employees to work on an OSS dependency. For example, P11 shares how in their company there are \inlinequote{dozens of people who contribute to [dependency name] on a pretty regular basis.} Similarly, P10 explains \inlinequote{I had about between 35 and 50, people whose full-time job  was to work on open source projects that the company used.}
%and engineers whose \inlinequote{full-time job was to work on open source projects (P10).}%that the company used [P10]}.

%With the time investment that companies put into their dependencies

With the time invested in OSS dependencies, companies aspire their \inlinequote{contributors to eventually try to move into maintainer positions, approver positions, leadership (P11)} (P3, P8, P13, P15) (see Figure \ref{fig:contribute}). This is particularly relevant for critical dependencies where \inlinequote{getting elected to leadership positions in these communities is really important (P3)} and allows companies to \inlinequote{help shepherd these projects forward (P18).}
Our results show that by investing developers' time in OSS dependencies, companies are sustaining the very projects that they depend on. This in turn (1) provides a business advantage and increases the company's reputation as one that \inlinequote{believe[s] there's a contribution wheel (P8)} and (2) benefits the OSS dependencies by maintaining a consistent contribution flow (see Figure \ref{fig:contribute}).

\subsubsection{Community Relations' Time}

Participants revealed that companies realize that it is necessary to build a relationship with the community to work in line with OSS, as mentioned by P9: \inlinequote{there are no special benefits to things in the open...if you're not able to build in the community.} This realization is at the core of companies investing time in community relations in addition to investing developers' time.

In this context, our results show that companies connect with OSS communities by holding open discussions (P5-P8, P15) which benefit both parties. P5 reported that, by doing this, \inlinequote{[the communities] learn more about what we're doing} and P15 complements it by saying that \inlinequote{[the company learns] what is it that people want.} These discussions include \inlinequote{talking about problems (P5),} \inlinequote{answer[ing] questions (P6),} and %when conflicts arise,
being open to compromise by \inlinequote{explaining the benefits, try to reducing the risks or problems of that solution, and trying to reach a better one (P7).} 

Some companies (P1, P11, P20) go the extra mile to help
foster community dynamics by organizing \inlinequote{contributor summits (P11)} and \inlinequote{events, where people [contributors] can come together (P20).} According to P20, this helps contributors to \inlinequote{meet each other.}% In addition, P11 mentioned that these events help the community attend \inlinequote{various community facing functions (P11).} 

Our findings suggest that contributing time to OSS is not only about contributing developers' time, but also spending time to nurture the community. P20 summarizes this as: \inlinequote{not just people spending time on code, but also people spending time on people}  which further magnifies the benefit to the project community (see Figure \ref{fig:contribute}).

%---------------ADVOCACY AND PROMOTION-----------------------
%Motivation: Reputation + recipro
\subsubsection{Advocacy \& Promotion Time} Our participants indicated that several OSS projects are tied to the software itself, making a lot of companies' contributions code related. But company participation does not stop at that. As our participants mentioned, they also \inlinequote{contribute to the overall ecosystem by doing advocacy work (P13)} and promote the prevalence of OSS (P5-P8, P11-P16).

%[general advocacy]
According to our respondents, companies advocate for OSS by encouraging participation internally and externally. Internally, companies encourage their employees to contribute to a project of their choice in their free time and compensate such contribution. P5 explains: \inlinequote{we encouraged our staff to contribute to an open source project that mattered to them, and then submit it for reimbursement.}
Externally, not only do companies \inlinequote{bring to the community, the awareness of how much this [open source] is important (P6)} but they also promote OSS to other companies. This takes the form of leading by example and as P6 describes:  \inlinequote{show to other companies in the ecosystem, that even if you're not a product company, you still may be able to invest money, time, people to contribute.}

%and explaining \inlinequote{how can they [companies] be part of the open source community (P14).} 

According to our participants, companies also encourage people to contribute by creating and sharing the resources to help them do so. During the interviews, they mentioned that the creation of resources takes the form of \inlinequote{blog posts (P6)} (P5, P7, P8, P16), \inlinequote{podcasts (P7)} (P5), \inlinequote{documentation (P15)} (P6-P9, P11), creating content for \inlinequote{open source month (P16),} creating and sharing survey results (P12), and participating in OSS working groups to codify best practices (P1, P18). 

We found that, while companies advocate OSS through attending existing conferences and events (P5, P7-P9, P11-P14), some take advocacy a step further by launching their own conferences, events, or hackathons (P1, P6, P7, P9, P11, P12, P20). These events help attract new OSS players by, for example, launching hackathons to help initiate newcomers to \inlinequote{solving some social issues (P6)} or promoting an OSS technology by \inlinequote{starting a conference... talking about open sourcing cloud (P7).}

%Companies also help nurture their existent contributor base by organizing \inlinequote{contributor summits [P11]} to help with contributors' experience and \inlinequote{various community facing functions [P11]} which further magnifies the benefit to the project community (see Figure \ref{fig:benefit}). 

%Developer/leadership time ---Management 
%Relation time (dev + community)
%mentoring time
%advocacy/ promotion time
%Addressing D\&I issues

%-----------------END OF ADVOCACY AND PROMOTION-----------------

%---------------------ADDRESSING D\&I------------------------
\subsubsection{Time Addressing D\&I} 
The participants (P5, P6, P13, P17, P19, P20) mentioned that their companies invest time and effort in addressing OSS Diversity and Inclusion (D\&I) issues. 

We could observe that the companies the interviewees work for contribute to the overall welfare of the OSS ecosystem (see Figure \ref{fig:contribute}) by investing effort in  \inlinequote{understanding how to be inclusive (P17),} understanding \inlinequote{diversity, equity inclusion initiatives (P20)} and understanding the ins and outs of the \inlinequote{code of conduct (P17).}

Once companies are well aware of the meaning and stakes of D\&I, they start thinking about different perspectives: \inlinequote{how to have these sorts of conversations (P20),} \inlinequote{how to support someone who is in, you know, feeling threatened, or unwelcome (P17)} and \inlinequote{how to have recognition and allyship (P20).}

%Who's missing? Why are they missing? Who's not talking? Why aren't they talking?[P17]

Only after building an understanding, do the companies start addressing D\&I issues. For some companies, this takes the form of \inlinequote{investing in more programs and initiatives that we hope will help just to improve that aspect of things [D\&I] (P20).} For other companies, it is more specific, focusing on geolocation diversity in OSS by contributing to OSS in their country of origin (P6). P5 shared that their company contributes to D\&I in OSS by partnering \inlinequote{with an organization that helps people learn how to develop. They come from an underserved population} and 
hiring \inlinequote{developers in countries that are less developed.} %and %, maybe they've never gone to college, or maybe they're doing a career transition [P5]}. 

%Interventions like these can help enrich the OSS ecosystem by bringing new faces in, and with that comes \inlinequote{more ideas from different cultures' perspectives and that also enrich[es] this innovation [P19]} thus closing the circle on the \inlinequote{virtuous cycle [P5]}.

%----------------END OF ADDRESSING D\&I------------------------

%--------------------MENTORING---------------------------
\subsubsection{Mentoring Time} Companies contribute to OSS by participating in mentoring initiatives, whether it be in a formal or informal capacity. By doing so, companies help attract and retain different OSS players, from individuals to other companies. This, in return, benefits the whole OSS ecosystem and ensures the continuity of different mentoring programs (see Figure \ref{fig:contribute}).

When it comes to informal mentoring, our findings show that companies give back to OSS by \inlinequote{mentoring people and always inviting others to contribute (P1)} (P14, P17). This is not only limited to individual contributors (e.g., newcomers) but also extends to helping companies that are not necessarily in the technical industry (e.g., ``manufacturing organizations,'' ``hospitals'' (P14)). According to P14, this is important since these companies \inlinequote{are struggling trying to figure out how to do this [open source].}

Whereas in formal mentoring, some companies run internship programs both internally and externally such as Google Summer of Code or Google Season of Docs (a mentorship program that nurtures technical writers). Other companies mentor aspiring contributors through these very same mentorship programs (P2, P9, P12). For instance, P9 mentioned
 \inlinequote{working with specific programs, like Google Summer of Code programs.}

At times, these programs can attract contributors and potential hires. For instance, P9's experience with mentorship programs resulted in \inlinequote{candidacies of folks that are willing to work with us.} Or, in the case of P2, \inlinequote{some contributors coming from Google Summer of Code (P2).} 

%and providing \inlinequote{overall positive force for the projects (P20).}
%-------------------END that donG---------------------

%---------------FINANCIAL SUPPORT-----------------------
\subsubsection{Financial Support} According to our interviewees, the companies participate in OSS by supporting one or multiple entities in the open source ecosystem. This can take the form of money set aside to support maintainers, foundations, patronage programs, and sponsoring events.

%Not only do companies \inlinequote{set aside time (P18)} but also \inlinequote{money to work with open source (P18)} (P16, P20).
Our participants mentioned that companies participate in OSS financially \inlinequote{maintaining the health of overall open source ecosystems (P15)} by: contributing money to dependencies (P4, P8, P5), hosting patronage programs (P4, P5), supporting maintainers (P5, P12), sponsoring events (P5, P11, P12, P15), foundations (P11, P12, P15, P16) and mentorship programs (P12) (see Figure \ref{fig:contribute}). 

Similar to investing time on projects that companies depend on, companies are also \inlinequote{more likely to patron that project if we're using it (P5).} This takes the form of \inlinequote{set[ting] aside a certain amount (P5)} (P8) to help these projects \inlinequote{pay the bills, and just like dedicate time for coding (P4).} A more specific way that companies invest money in dependencies is by directly supporting the maintainers (P5, P12, P15) through, for instance, \inlinequote{hiring open source maintainers full time (P12).} 

While providing direct financial support to OSS dependencies helps ensure the health and dependability of specific projects, companies also believe in supporting OSS foundations. For instance, P12 explained that \inlinequote{if we don't also think about supporting the foundations that provide a home for those projects, then we're not really going to solve the problem in its entirety.} Our results show that companies \inlinequote{give a lot of money to foundations (P11)} (P12, P15) to \inlinequote{pay for infrastructure (P15)} and \inlinequote{to pay people to do all the things that they do...to support the open source projects (P11).} This opens the door for companies to be involved in \inlinequote{safety and security (P8)} conversations that are usually \inlinequote{exclusive to sponsored members (P8).} 
%%%%%%
\subsection{Tying it together: A summary}
Having described the motivations that lead companies to contribute to OSS and the different ways companies contribute to OSS, we now tie everything together, summarize our findings, and discuss lessons learned from our interviewees on OSS sportsmanship and being a good OSS citizen.

With 97\% of companies using and relying on OSS \cite{ossCompanyPercent}, profit aside, companies' well-being is now tied to OSS' well-being where the investment in OSS is not just linked to a short-term profit, but rather an investment in the future. Companies invest in reciprocal contributions such as mentoring and advocacy and promotion, as well as addressing D\&I issues, and go as far as archiving OSS code in the Arctic for future generations~\cite{githubArcticOSS}.

\begin{figure}[hbtp]

    \includegraphics[width=.49\textwidth]{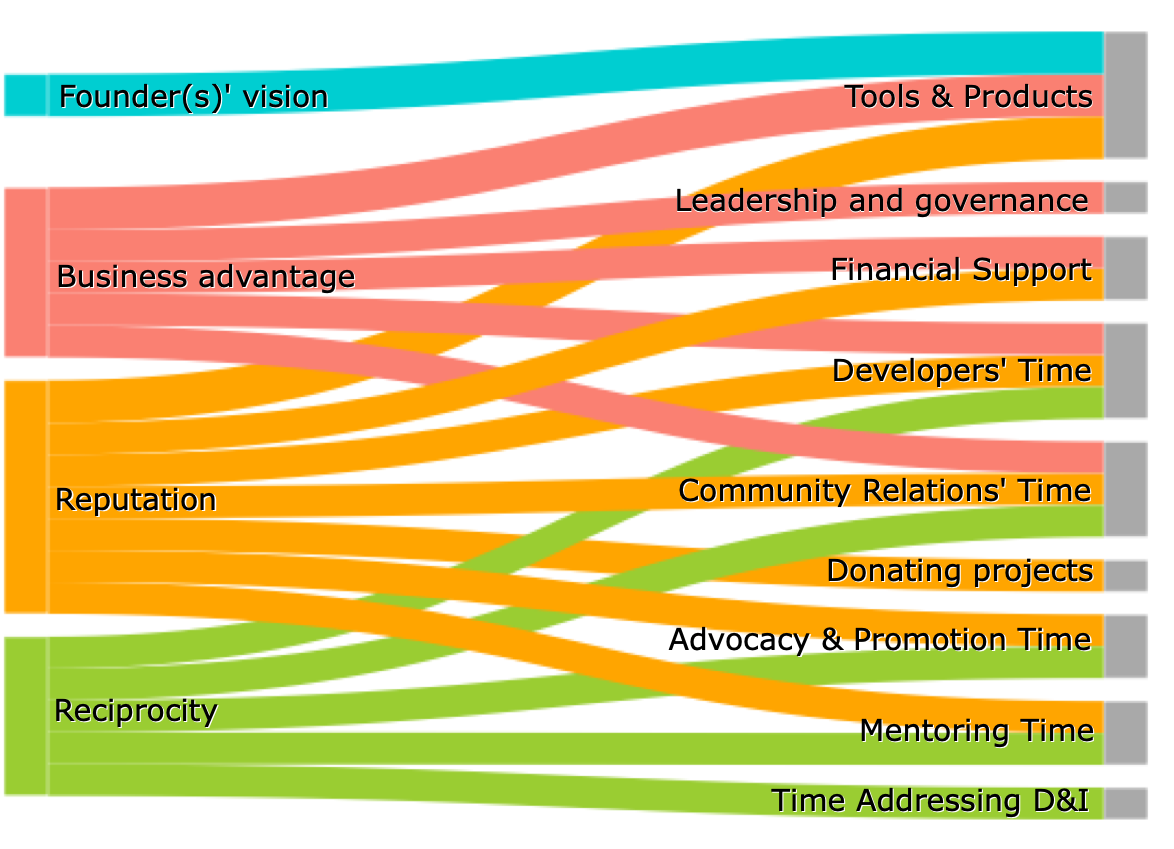}
    %\vspace{-3em}
    \caption{Relationship between motivations and the type of contributions}
    \label{fig:motivation}
    %\vspace{-0.8em}
\end{figure}

Companies are realizing that OSS is also about innovation, being part of the collective brain power, being a humble participant, and doing good for the community.
Company participation in the OSS ecosystem is multi-faceted, beneficial to different entities (see Figure \ref{fig:contribute}), and intertwined with different motivations (see Figure \ref{fig:motivation}). Figure \ref{fig:motivation} depicts how the different types of company contributions (right-hand side) flow from the different motivations (left-hand side).

Increasing reputation by contributing to OSS was one of the most prevalent motivations. Companies increase their reputation not only by open-sourcing tools and products and investing their developers' time but also gain visibility by investing in non-code related activities (e.g., advocacy \& promotion, mentoring).

While companies also contribute time (developers' and community relations), financial support, and software to maintain their business advantage, being a good OSS citizen is a pertinent driver for a variety of contributions. In fact, companies see the benefit in being a humble OSS participant where they do not just take from, but also give back to the community (see reciprocity, Figure \ref{fig:motivation}). One key aspect was taking the time to address issues of Diversity and Inclusion (D\&I). D\&I is an important topic in open source yet a challenging one to address. When companies, that have the tools, resources, and experience, take the initiative to champion D\&I in OSS, it is a significant step forward. 

%While companies recognize the impact of OSS on their business and profit, 
Being a good OSS citizen (see Table \ref{table:sportsmanshipTable}) was seen as key to healthy OSS participation. One way to do so is by \inlinequote{contribut[ing] in the way that makes sense to the project, which may or may not always be, again, what the company would want (P20).} This entails placing the community's needs first and helping out with maintenance tasks.
%, and to parts of the project that they are benefiting from. 

In order to participate in an OSS ecosystem, an open source program in a company needs to be \inlinequote{able to speak multiple languages (P10)}: risk management, engineering benefits, legal \& compliance, and sponsorship, to be able to advocate OSS to \inlinequote{different levels of leadership. And showing how open source helps the business (P18).} 

%Another, is by \inlinequote{tak[ing] that technical depth task that clean out things, that restructure things (P7)} and lessening maintainers burnout.
Finally, it is good to have an abandonware strategy. To create a sustainable project, companies need to open-source projects they use and care about internally. It is also important to have a contingency plan if the project does go out of sync internally by \inlinequote{transfer[ing] it to people who care about them [abandoned project] (P4).} Ensuring that the project does not become a company monopoly and has affiliation diversity is a good step in avoiding abandonment crisis.

\begin{table*}[bht]
%\footnotesize
%\scriptsize
\caption{OSS Sportsmanship: lessons learned}
\vspace{-1.5mm}
\begin{tabular}{p{0.2\textwidth}p{0.1\textwidth}p{0.6\textwidth}}
\toprule
{\textbf{Takeaway}} & {\textbf{Participants}} & {\textbf{Exemplary Quotes}} \\ \hline

%\textbf{Play by the community rules:} The community needs take precedence over the company's needs 
%\newline
\textbf{Don't bulldoze your way into getting what you want:} The community needs take precedence over the company's needs 
&

 P5-P8, P11, P14-P20
&
``You really have to think of the community first...we can't just bulldoze our way into getting what we want. We have to think about how what we want benefits the community'' (P11)
\newline
``We try to pick the things that would be having a bigger impact, bigger benefit for the whole of the community'' (P7) %... The main chunk of the work we do is what can we do for the community that will help them'' 
%\newline
%[This is mentioned in the companies code of business conduct] 
%``So if (a) is right for the project, because that's what you need to do, even though it's negative to [company name], yeah, it doesn't matter. It's important that the project succeeds'' [P14]
\\

\cellcolor{gray!15}\textbf{Prevent maintainer burnout:} Participate in maintenance tasks 
& 
\cellcolor{gray!15} P7, P17, P18, P20
&

\cellcolor{gray!15} ``Something needs to be shiny, for you to do it on your free time...companies that have the means to provide people to work on open source should be doing those things'' (P7) 
%\newline
%``learning how you can help and we often talk about that chopping wood, and carrying water thing... sometimes the most useful things you can do are the non-glamorous things... humble participation is most important'' [P17]
\\

\textbf{Keep the contribution wheel spinning:} Where you consume, contribute 
& 
P4-P9, P12, P20
& 
``Everyone should be contributing back, everyone should be making sure that they are not just the receiving end, they are receiving and producing end of the open source'' (P7)

``We generally believe there's a contribution wheel, so we owe, if we can upstream something or contribute back to a project we use, then we will'' (P8) 
\\

\cellcolor{gray!15}\textbf{Prevent corporate abandonment:} Open source projects that your team cares about
& 
\cellcolor{gray!15} P4, P17
& 
\cellcolor{gray!15} ``We do want to open source things that play into our own brand of the company...If it's something that is a core part of our teams' function, that's also great because it's something that they [we] will continue working on'' (P4)
%\newline
%``it looks terrible when there's like abandoned repos, if a company just puts something... looking at the upstream, making sure that your engineers are not just, you know, we're very mindful'' [P17]
\\

%2:26 ¶ 151 in [P4] Per-Spotify interview.docx
%e is some some sort of reason why you actually open sourced it, beyond just like personal growth, and actually becomes an integrated part into, like, why is the team considered successful? If it's not, then it's going to be left by the wayside at some point, because they'll be more important things. But if it's actually an integral part of the, the objectives of the team, then they will also continue to be maintained.

%And then you could say on the other end of the scale is like things we do want to open source is like, especially things that play into our own brand of company. Like if it's something about music and technology that would be amazing. If it's something that's a core part of our teams function, that's also great because then it's something that they will continue working on[P4]

%just know what you why you're doing it, you know, just coming out and be like, Oh, we open source something. So that's really important. And then like, it looks terrible when there's like abandoned repos, if a company just puts something else that's around release, around use, again, is like, yeah, looking at the upstream, making sure that your engineers are not just, you know, we're very mindful, a lot of times we have compliance software to help them be mindful.[P17]

%-------------------------------------------------------
\textbf{Have an abandonware strategy:} Transfer ownership of OSS projects that you internally moved away from but the community cares about
& 
P4, P6, P10, P15, P17
& 
``If we stopped caring about them [open sourced projects that the company moved away from], then we should transfer it to people who care about them'' (P4)
\newline
%``We have a project that is actually a big success, like open source wise, but we don't use it internally anymore, and we don't really care about anymore... But outside of [company name], it is, it's a perfectly fine living project with a lot of different contributors'' [P4]
%\newline
``Eventually, you stop using the tool internally. But you know, have partners and customers that are using it...you can just kind of gradually pull all your engineers off of it, and let them have it. Um, you know, so this is what's called the abandonware strategy'' (P15)
%... where you say, ``Hey, I know you guys are still using this. I know you still like it, but it doesn't make any sense for us as a business anymore. So here, it's yours. Now you can have it'' [P15] 
\\

\cellcolor{gray!15}\textbf{Avoid company monopoly:} Invite other companies in 
& 
\cellcolor{gray!15} P3, P5, P7, P11, P15, P17
& 
\cellcolor{gray!15} ``Part of the healthiness of the project is also making sure that there are different ideas, opinions, and points of view, it's really easy to fall to a kind of monopolistic way'' (P7) % the best things of the open source is that it's developed in the open with everyone'' 
%\newline
%``For a very successful public open source project, that is one where you have a public community of people contributing and involved, who work for all kinds of different employers, or no employer'' [P15]% or, you know, all over the place'' 
\\

 \textbf{Be a polyglot in promoting OSS:} 
Speak risk management, engineering benefits, legal \& compliance, and sponsorship 
& 
 P7, P10, P14, P16-P19
& 
 ``An interesting challenge for a successful open source program in a company is to be able to speak multiple languages, to speak to the risk legal people about risk strategy and, you know, compliance, to speak to the engineers about the benefit of being in the goodwill...and then to speak to the CFO, about, here's how we save money, here's how we make money'' (P10) %we save money on these proprietary licenses, and on the tech debt... and all those are the three conversations, the fear, risk management, the love, community, goodwill, and the money are all part of the conversation. So an OSPO has to negotiate between'' [P10]
%\newline
%``The OSPO will be successful, if they manage to convince or to make sure that the philosophy of open source is spread across the company'' [P7]
%\newline
%``there are people who've never worked in open source before and who...are learning or might be fearful. So, I think that's the big picture, but that there are lots of little individual feelings, stories, people have different levels of learning'' [P17] 
\\

\bottomrule
\end{tabular}
\label{table:sportsmanshipTable}
\vspace{-2.5mm}
\end{table*}

\section{Threats to validity}
\label{sec:threat}
 %it's a threat but this is what we did to minimize it...
 %-saturation?

\textbf{Construct validity} in qualitative research is related to the definition of constructs. One issue could arise from asking incorrect questions which we mitigated by piloting the interview script with the research team. Another issue pertains to the qualitative coding process. To mitigate this and avoid misinterpretation, we compared new emerging codes with the existing code set \cite{barney2017discovery} and met frequently with the research team to discuss and clarify the codes. Additionally, the final code set was reviewed and finalized by the research team. To maximize credibility, we depict examples of evidence from observations to findings (see Table \ref{tab:motivationTable}).

%During the coding phase, we used the
%constant comparison technique \cite{barney2017discovery} whereby each interpretation and finding is compared with
%existing findings as it emerges from the data analysis to increase the construct validity of our findings.

\textbf{Internal validity} is related to capturing reality as closely as possible, which in our case is company motivations and contributions to OSS. The characteristics of our sample may have influenced our results. Half of our interviewees were part of the OSPO team within their company even though we did not push for such a split. The sizes of companies from which we interviewed participants were balanced (7 small, 1 medium, 9 large). We reached saturation after the $13^{th}$ interview. We also performed member checking to validate our constructs and received validation from all eight of our respondents.

% The interviewed participants were evenly distributed across different-sized companies  (7 small, 1 medium, 9 large)
% 12/ 17
% balanced company size 

%To accurately capture these motivations and contributions, our sample includes participants from different-sized companies who have different roles. Our participants' roles are related to some OSS part of their company, which validates the knowledge they contributed to our study. Additionally, we validated our constructs via member checking where we received validation from all XX of our respondents.  

\textbf{Reliability} refers to the extent that our results can be replicated. In short, it is difficult to replicate qualitative research since human behaviors, feelings, and perceptions change over time. However, we maintained consistency by constantly comparing the analysis with already existing codes and having weekly meetings to discuss and adjust codes and categories until
we reached an agreement. We also performed member
checking with eight participants who confirmed our interpretations. Finally, the results presented in this paper are related to Open Source participation, thus, we do
not expect that all our findings will apply to other contexts.

\textbf{Theoretical saturation.} A potential threat to validity regards reaching theoretical saturation. The quality, rather than the size, of the sample of participants, is essential to increase our confidence in the findings. We interviewed 20 participants with different OSS-related roles. %and the other half were part of non-OSPO related roles that are knowledgeable about the companies' participation in OSS (e.g., CEO, Ecosystem Strategist). 
The 20 participants represented 17 companies of different sizes, from big technology companies, such as Microsoft and Google, to growing startups. We kept interviewing participants until no new codes emerged (13 interviews). Moreover, we conducted seven additional interviews and we did not find any new constructs. %As mentioned previously, the number of interviewed participants was adequate to uncover and understand the core categories in any all-defined cultural domain or study of
%lived experience \cite{}. 
Although we cannot claim theoretical saturation, our sample helped us uncover a consistent and comprehensive account of the nature of company motivations to contribute to OSS and uncover the different forms of contribution.

\section{Related Work}

%%%%_____________Mariam Comments_____________________________
%Most oss work focused on individual/project experiences in oss from their motivation to challenges 

%When it comes to companies in OSS, research focuses mainly on innersource, ways to apply OSS processes to a company [] benefits of innersource [] and business models where in that case they .....

%However, there is little to no understanding of why and how companies get involved in open source.... 

%%%%_____________Mariam Comments______________________________

\boldification{Handful of literature on individual's participation in open source}
%Motivation, challenges, barriers, newcomers, experienced contributors...
\boldification{innersource--just mentioning that there is innersource out there, as well as OSPOs, but we're not focusing on that}
%some people explored hybrid setting ... 
%the ones that explore how companies get involved we go more in-depth 
%Existing literature on companies' involvement in open source spans a range of focus, from innersource \cite{} to companies' OSPOs \cite{}. Additionally, 

A vast majority of prior literature in OSS has focused on individuals' experiences, from their motivations to participate \cite{gerosa2021motivation, lee2017understanding, von2012carrots, hannebauer2016motivation, oreg2008exploring , roberts2006understanding, huang2021leaving, yang2022projects} to challenges faced when participating \cite{steinmacher2015systematic, steinmacher2015social, guizani2021long ,lee2017understanding, huang2021leaving} and their perception of the OSS ecosystem \cite{lee2019floss, guizani2022perceptions, vasilescu2015perceptions}.

\textbf{Differences between individual and company motivations.}
Individuals' motivations to contribute to OSS has been extensively studied \cite{gerosa2021motivation, lee2017understanding, von2012carrots, hannebauer2016motivation, oreg2008exploring , roberts2006understanding, huang2021leaving}, with the literature review by Von Krogh et al.  \cite{von2012carrots} being the most comprehensive and Gerosa et al. \cite{gerosa2021motivation} work, which builds on \cite{von2012carrots} being the most recent. %In this section we draw a parallel between our findings on companies motivations and prior literature's findings on individuals' motivations. 

From all the individual ``intrinsic'' motivations (Ideology, Altruism, Kindship, Fun), only \textit{Ideology} has a company equivalent. \textit{Founder(s)' vision} is an individual's \textit{Ideology} turned corporate. \textit{Founder(s)' Vision} as a motivator for participation was identified for small companies where the founder(s)' beliefs carried onto their company, defining its culture and DNA.

All individuals' ``internalized extrinsic'' motivations (Reciprocity, Reputation, Own use), except Learning, have their parallel in companies. \textit{Reciprocity} is a common motivation for both individuals and companies, where both actors aim to give back to the community to reciprocate the benefits they got from OSS.
Individuals participate in OSS to improve their own \textit{Reputation} \cite{gerosa2021motivation} and companies their brands. For an individual, gaining reputation can be a segue to getting hired by companies who themselves participate in OSS to attract talent. Companies benefit by hiring individuals who have been working on the project allowing companies to bypass recruitment and onboarding costs. Companies' \textit{business advantage} maps to individual's \textit{own use} but adds a more nuanced definition, since %While \textit{own use} by definition maps to companies' \textit{engineering need}, 
\textit{business advantage} also includes drivers such as coopetition, having closer channels, and improved innovation.

Individuals' ``extrinsic motivation'' (getting paid, career) do not directly map to companies. Companies do not get paid (or have contracts) to contribute to OSS, instead, they are the ones who may sponsor OSS projects. However, companies do indirectly profit from OSS participation as a result of an improved reputation and recruiting talented individuals.

\textbf{Company participation in OSS.}
%innersource + legal compliance
When it comes to companies adopting OSS processes, prior literature investigated innersource adoption~\cite{stol2014key, stol2014inner, edison2020inner, morgan2011exploring, capraro2018patch, carroll2017value}, and the motivations and challenges of applying it~\cite{capraro2016inner, carroll2018examining, gaughan2009examination, stol2011comparative}. Researchers also identified compliance best practices to help companies avoid legal/IP risks when using OSS~\cite{harutyunyan2019getting, harutyunyan2019industry, harutyunyan2018understanding}. OSS-related business models used by companies have been analyzed and compared as a way to understand the benefits and risks of each model~\cite{munga2009adoption, spijkerman2018open, li2019does, mouakhar2017open, deodhar2012strategies}. Similarly, the literature showed how companies' usage of OSS relates to their business models~\cite{dahlander2008firms, riehle2011controlling, morgan2014beyond}. Companies use OSS to increase their productivity and product quality~\cite{ajila2007empirical}. This, in return, encourages companies that started with closed in-house software \cite{andersen2012commercial} or hardware~\cite{li2021understanding} to open source them by following a specific pathway \cite{kochhar2019moving, mortara2011large, zynga2018making, pruett2013comparison}. This results in more hybrid communities, where the intensity of companies' involvement helps understand their impact on volunteer communities \cite{zhang2019companies, zhou2016inflow, dahlander2005relationships, aagerfalk2008outsourcing, maenpaa2018organizing}.

% ways to contribute 
%Furthermore, researchers have identified some work practices companies use when contributing to a selection of five established OSS projects, some of which  include ``employing core project developers, and joining project steering committees in order to advance strategic interests'' \cite{butler2018investigation, butler2019company}. 

%the communication practices between companies and foundation-owned OSS projects when making a contribution (e.g., an explanatory question on a mailing list when reporting a bug) and the reason behind using public channels
%(e.g., mailing lists, changelogs) \cite{butler2018investigation, butler2019company}. While Butler et al.'s looked at how companies communicate with OSS projects when they make a contribute, we looked at why companies contribute in general and what types of contribution they engage in. 

%_________compare with results___________
Butler et al.~\cite{butler2019company} investigated the communication practices between companies and foundation-owned OSS projects (e.g., questions on a mailing list when reporting a bug) and the reason behind using public channels (e.g., mailing lists, changelogs). They identified the ways for companies to communicate with OSS projects and reasons for making it public. While Butler et al. focused on companies' communication practices when making a contribution, we focused on the multi-faceted ways companies contribute to OSS and identified all the kinds of ties that companies create with OSS and the motivations that lead the company to join OSS. Still, in their case study, Linaker and Regnell found 12 objectives for deciding what software to open source and when. While Linaker and Regnell \cite{linaaker2020share} work focused on a specific type of contribution (open-sourcing software) we focused on a broader perspective, aiming to build a comprehensive motivation model that drives all kinds of contributions, their ties to the motivations and the entities that they benefit. 

Our work complements this body of research by focusing on companies as OSS contributors instead of looking at individuals. We also identify the ways to engage and the reasons that lead companies to engage with OSS. We contributed a comprehensive model of company motivations to contribute to OSS beyond open-sourcing products. Indeed, we show that the ways to contribute are multi-faceted, are tied to the motivation of the companies, and may influence different entities.

%97\% of companies use open source (GitHub talk)
%Linux foundation surveys

\section{Concluding Remarks}
\label{sec:discussion}
\boldification{reiterate the importance of this work--similar to part of intro}
Understanding the \textit{why} and \textit{how} companies contribute to OSS helps create a more comprehensive picture of the different players in the OSS ecosystem. 
Our results reveal that participating in OSS helps companies gain a business advantage and also grow their reputation. Reciprocating by giving back to the community is another key motivator. Smaller and larger companies often differ in why they participate in OSS. For example, when considering \textit{reputation} as a motivation, smaller companies participated as it allowed them to gain reputation and to participate in the same space as big-name companies, whereas larger companies reported that OSS engagement signals good citizenship helping cement their reputation in the community.

%\textbf{Lessons learned on OSS sportsmanship}
%Recurring sentiments about participation in OSS appeared throughout our interviews that reflected virtuous ``OSS sportsmanship'' (see Table \ref{table:sportsmanshipTable}). Here, we highlight a few lessons in prose from fellow companies on the OSS cause: 
%On companies' sportsmanship, \textsc{don't ``bulldoze'' your way into getting what you want} and \inlinequote{contribute in the way that makes sense to the project, which may or may not always be, again, what the company would want (P20).} \textsc{But take the less shiny tasks and help prevent maintainers' burnout}-- \inlinequote{tak[ing] that technical depth task that clean out things, that restructure things (P7).} 
%When it comes to open sourcing a project,  \textsc{prevent abandonware} by open sourcing projects your company uses internally and cares about. \textsc{Though if it happens} and the project goes out of sync \textsc{have a strategic call} by  \inlinequote{transfer[ing] it to people who care about them [abandoned project] (P4).} 
%\textsc{And finally, be a polyglot}: \inlinequote{be able to speak multiple languages (P10)}--fear, risk management, community, and money when advocating OSS to \inlinequote{different levels of leadership. And showing how open source helps the business (P18).} 

\textbf{The OSS renaissance is well underway.}
Open Source has its roots in the Free Software movement, which emerged as the antithesis of the commercial software movement. Without going into the discussion comparing these movements, it is possible to notice that OSS grew and expanded far beyond the weekend warriors. %and the fight against corporations. 
OSS is no longer produced and maintained by a dedicated few, but rather now is a space for different players of different capacities to thrive together--triggering the positive feedback loop of technology and business innovation. 

While tech companies are invested in OSS now more than ever, non-tech and low-tech industries are also seeing the value of OSS (e.g., financial companies) and are establishing Open Source Program Offices (OSPOs) to systematize their contributions and policy implementations. Universities (e.g., John Hopkins, and Rochester Institute of Technology) are also participating by investing in OSPOs to centralize their OSS activities and simplify inter-university and company-university collaboration. %away from the contracts, licenses and grant hustles. 
Organizations such as UNICEF and the United Nations are looking to work with the OSS community to help reach their sustainable development goals  \cite{opensourceUN, sdgsUN}. With a major part of national infrastructures being dependent on OSS (e.g., toll booths, hospitals), municipalities such as the City of Paris have their own OSPO and are contributing city-level digital solutions (e.g., Lutece~\cite{luteceParis, luteceGitHub}) and it is only a matter of time before we start seeing nationwide efforts to invest in OSS and government OSPOs. 

\textbf{Implications for companies.}
Our findings can help nudge companies to contribute to OSS by understanding the benefits of OSS for their business, their branding, and the overall OSS ecosystem. Companies can then identify the motivations that are important to them and prioritize their contributions accordingly. Our mapping between the motivations, ways to contribute, and the benefiting entities, can help companies be more systematic about their OSS presence. Not only that, but the lessons learned from our company participants can serve as best practices for engaging with the OSS community; practices that help foster a high level of sportsmanship and a symbiotic OSS-company relationship. It is important for companies to recognize that they may be a crucial part of the sustainability of the project, as many interviewees mentioned reciprocity as a motivation to join. Reciprocity not only works as a ``social responsibility'' mechanism but as a way to guarantee that the project will be healthy, which in turn benefits the company. We expect that our results enlighten companies, motivating more organizations to see the benefits of engaging in OSS.

\textbf{Implications for OSS projects.}
When company involvement in OSS is done right (Table \ref{table:sportsmanshipTable}), it creates a symbiotic ecosystem where both parties mutually benefit. OSS projects can use our findings to help attract company participation and solicit contributions for their project needs.

By understanding companies' motivations, projects can provide an environment that is attractive to companies. For instance, reputation is a driver for companies to contribute, so if the project's community is \inlinequote{an abusive community, even companies don’t want to get involved (P18).} This is an additional impetus for projects to foster a healthy, respectful community to attract company participation. To retain these companies, the project can provide open leadership and governance, one where everyone is \inlinequote{welcome to provide input on the governance and directions of the project (P9).} This can nudge companies to depend on and contribute to a project in a more systematic way, \inlinequote{know[ing] that they're contributing on a level playing field (P11).}

\textbf{Implications for researchers.}
Interventions designed to help sustain OSS need to take into account the different OSS players, their goals, and how they contribute. Our study complements the expansive literature on individuals' experiences in OSS by investigating companies as players. Researchers can investigate other OSS players (e.g., Universities, Government bodies), the interplay of these players, or the role OS    S can play in facilitating collaboration between these parties. Looking at the other side, it is also important to understand the perspective of individual projects and contributors when they are a part of a company-owned or company-sponsored ecosystem, and how the motivations from the different sides align. 

In conclusion, we hope our results encourage and provide guidance for current and new participants in the OSS ecosystem--specifically companies--to engage, collaborate in the open, create better software, and broaden the economic pie. 
%We believe that  lessons from this study will help companies and OSS projects continue to foster this symbiotic relationship, where each sustains the other effectively.

%the company goals and ways to contribute as one of the players. 

%\input{conclusion.tex}
%\input{Acknowledgment.tex}

% BALANCE COLUMNS
%\balance{}
\bibliographystyle{IEEEtran}
\bibliography{biblio}
\end{document}